%Paper: astro-ph/9404066
%From: white@merkin.astr.ua.edu (Raymond E. White III)
%Date: Tue, 26 Apr 94 13:18:51 -0700

\magnification=\magstep1
\hsize=6.5truein
\vsize=9.truein
\hoffset=-0.2truein
\voffset=-0.12truein
\parskip=0pt
\parindent=25pt
\baselineskip=21pt
\font\tensmc=cmcsc10
\def\smc{\tensmc}
\def\br{{\hfil\break}}
\def\etal{{\it \etal\ }}

\def\mdot{{\dot M}}

\def\msun{M_\odot}
\def\msunyr{\msun~{\rm yr}^{-1}}
\def\spose#1{\hbox to 0pt{#1\hss}}
\def\lta{\mathrel{\spose{\lower 3pt\hbox{$\sim$}}
	\raise 2.0pt\hbox{$<$}}}
\def\gta{\mathrel{\spose{\lower 3pt\hbox{$\sim$}}
	\raise 2.0pt\hbox{$>$}}}
					\baselineskip=12pt
					\null
					\vskip 1 truein
					\centerline{\bf
ABUNDANCE GRADIENTS IN COOLING FLOW CLUSTERS:	}
					\centerline{\bf
$GINGA$ LAC \& $EINSTEIN$ SSS SPECTRA OF	}
					\centerline{\bf
A496, A1795, A2142 \& A2199		}
					\vskip 0.5 truein
					\centerline{\smc
Raymond E. White III$^1$		}
					\centerline{\smc
C. S. R. Day$^{2,3}$			}
					\centerline{\smc
Isamu Hatsukade$^4$			}
					\centerline{\smc
and					}
					\centerline{\smc
John P. Hughes$^5$			}
					\vfil
					\noindent{
$^1$Department of Physics and Astronomy, University of Alabama, Box 870324,
Tuscaloosa,}				\br
					\noindent{
$^{\ }$\ AL 35487-0324	}\br
					\noindent{
$^2$Laboratory of High Energy Astrophysics, NASA/GSFC, Greenbelt, MD 20771}
					\br
					\noindent{
$^3$USRA Research Associate		}\br
					\noindent{
$^4$Department of Electronic Engineering, Miyazaki University, 1-1 Gakuen
Kibanadai,				}\br
					\noindent{
$^{\ }$Nishi, Miyazaki, 889-21, Japan	}\br
					\noindent{
$^5$Harvard-Smithsonian Center for Astrophysics, 60 Garden Street, Cambridge,
MA 02138}\br
					\eject

					\centerline{
ABSTRACT				}
					\medskip

We analyze the $Ginga$ LAC and $Einstein$ SSS spectra of four cooling flow
clusters, A496, A1795, A2142 and A2199, each of which
shows firm evidence of a relatively cool component.
The inclusion of such cool spectral
components in joint fits of SSS and LAC data leads to somewhat higher
global temperatures than are derived from the high energy LAC data alone.
We find little evidence of cool emission outside the SSS field of view.
Metal abundances appear to be centrally enhanced in all four cluster, with
varying degrees of model dependence and statistical significance:
the evidence is statistically strongest for A496 and A2142, somewhat
weaker for A2199 and weakest for A1795.
We also explore the model-dependence in the amount of cold, X-ray absorbing
matter discovered in these clusters by White $et$ $al.$ (1991).
\smallskip
\noindent{{\it Subject headings:} galaxies: abundances ---
galaxies: clustering --- galaxies: intergalactic medium --- galaxies: X-rays
--- X-rays: general}

					\vfil\eject

					\bigskip
					\centerline{
1.\ \ INTRODUCTION			}
					\medskip

Some of the most accurate temperatures and iron abundances available for the
hot gas in many galaxy clusters have been obtained from X-ray spectra taken
with
the Large Area Counters (LAC) on the $Ginga$ satellite  (Takano $et$ $al.$
1989;
Hatsukade 1989; McHardy $et$ $al.$ 1990; Koyama $et$ $al.$ 1991;  Day $et$
$al.$
1991; Arnaud $et$ $al.$ 1991; Birkinshaw $et$ $al.$ 1991;  Ikebe $et$ $al.$
1992;
Allen $et$ $al.$ 1992;
Arnaud $et$ $al.$ 1992; Johnstone $et$ $al.$ 1992; Hughes and Tanaka 1992;
Hughes $et$ $al.$ 1993).
However, with an effective lower energy limit of $\sim1$ keV, the LAC was not
very sensitive to the spectral signatures of cooling flows which occur
in the centers of many clusters.
In contrast, the lower-energy $Einstein$ $Observatory$ Solid
State Spectrometer (SSS) and Focal Plane Crystal Spectrometer (FPCS)
were particularly sensitive to emission from cooling flows passing through
temperatures $kT\lta1$ keV
(Mushotzky and Szymkowiak 1988; Canizares, Markert and Donahue 1988).
Recent re-analysis of cluster SSS spectra has also revealed large
quantities of relatively cold, X-ray-absorbing
gas in many  cooling flow clusters, quantities well in excess of the
Galactic column of cold material in their lines of sight
(White $et$ $al.$ 1991, hereafter WFJMA).  This intrinsic absorption
is attributed to the accumulation of cooling flow condensates and may account
for much of the total material likely to have been accreted during the lives
of the cooling flows (WFJMA).  Wang \& Stocke (1993) also found evidence
of absorption in $Einstein$ IPC spectra of distant clusters.
Furthermore, spatially resolved $ROSAT$ PSPC spectra of the copious cooling
flow cluster A478 confirm the SSS evidence for an absorption component and also
show that it is indeed confined to the cooling flow (Allen $et$ $al.$ 1993).

While many clusters have had analyses of their SSS or LAC spectra published,
only one cluster, A478, has had its LAC and SSS data analyzed together,
although
the spectral models were not fit simultaneously (Johnstone $et$ $al.$ 1992).
Joint analyses of SSS
and LAC spectra are particularly informative because their different passbands
provide complementary constraints on spectral models, particularly when their
differing fields of view are taken into account: the SSS had a circular
field of view with a 6$^\prime$ diameter (flat top),
while the LAC had a much larger,
elliptical field of view, $1^\circ\times2^\circ$ (FWHM).
The complementary nature of these data is illustrated by considering a
cluster which is largely isothermal except for a central cooling flow which
is cooling from the same temperature as the exterior cluster gas.   If such a
cluster were nearby, a centered SSS spectrum would be dominated by the cooling
flow emission, while the LAC spectrum would include emission from the entire
cluster.   The cooling flow spectrum would be determined in part by its initial
temperature, but this temperature would tend to be underestimated by an
analysis
of the soft SSS spectrum alone.
The LAC spectrum, in turn, would provide an accurate global
temperature and abundance, but would not be able to constrain the properties of
the cooling flow (accretion rate, abundances, internal absorption) very well
because much of the cooling flow emission and the effects of internal
absorption would be most evident at energies lower than the LAC passband.

The power of joint spectral analysis is best realized by
fitting the complementary spectra simultaneously.  Hughes and Tanaka
(1992) simultaneously analyzed the $Einstein$ IPC and $Ginga$ LAC
spectra of A665 and were able to constrain its temperature distribution.
These authors found this could not have been achieved by
analyzing the data sets separately.  Moreover, simultaneous fitting allows
more accurate assessment of the parameter uncertainties, since all the relevant
model fitting parameters can co-vary in the course of error analysis.

Inspection of the isothermal fits to LAC spectra in Hatsukade (1989)
reveals persistent positive residuals below $\sim2$ keV in several
clusters known to have cooling flows.  Whether these positive soft residuals
are due to central cooling flows or, say, to cool emission in the outer parts
of clusters may be distinguished by a joint analysis of the SSS and LAC
spectra.
If the cool emission is fully accounted for within the smaller SSS
field-of-view, the cool emission seen in the LAC spectra is not likely to
come from cluster exteriors. The differing fields of view can also be
exploited to determine whether there are gradients in abundances or cool
X-ray absorbing material.

In this paper, we investigate these possibilities with a joint analysis of
the LAC and SSS X-ray spectra for four clusters, A496, A1795, A2142 and A2199.
In $\S2$ we describe how the cluster X-ray spectral data were reduced.
We analyze the four individual clusters in $\S3$, in order of Abell number
(which are in right ascension order). A2199 and A496 have the highest
quality data, followed by A1795, while A2142, the most distant cluster of
the four, has the poorest statistics.  We summarize and further discuss our
results in $\S4$.  Imaging data will be incorporated in later studies of
individual clusters.

					\bigskip
					\centerline{
2.\ \ DATA REDUCTION			}
					\medskip

%					\bigskip
					\centerline{\it
2.1\ \ Ginga LAC spectra			}
					\medskip

The $Ginga$ observations presented here of A496, A1795 and A2142 have been
analyzed before by Hatsukade (1989), who primarily investigated isothermal
models.  For these clusters, we use a similar set of reduced  $Ginga$ data as
Hatsukade (1989).  Analysis of the $Ginga$ observations of A2199 has not been
previously published.

The eight Large Area Counters (LAC) on  $Ginga$ covered the energy range
1--37 keV with 48 channels and an energy resolution of 18\% at 6 keV.
The (unimaged) field of view was $1^\circ\times 2^\circ$ FWHM and
the combined geometric collecting area was 4,000 cm$^2$.
Descriptions of $Ginga$ and the LAC can be found in Makino $et$ $al.$
(1987) and Turner $et$ $al.$ (1989), respectively. The data for A496, A1795
and A2142 are collected from the top and middle of the three LAC layers,
while the data for A2199 are from the top layer alone.  A log of the
$Ginga$ observations is given in Table 1. The $Ginga$ observations of these
four
clusters yielded spectra integrated over the entire cluster and their immediate
surroundings with high signal-to-noise but modest spectral resolution.

Of primary concern when reducing $Ginga$ data from moderate to faint sources is
the proper subtraction of the background, both X-ray and  non-X-ray. We adopt
the most straightforward way of removing the background: subtracting an
observation of a nearby  blank field taken on an adjacent day. By observing a
nearby field, the dependence of the diffuse X-ray background on sky position is
accounted for, while observing on an adjacent day ensures that the geomagnetic
conditions of the background closely match those of the cluster. In addition,
we discard data collected from the five (of fifteen) satellite orbits per day
which passed through the South Atlantic Anomaly, thereby incurring a high
background.

Each spectrum was associated with a response matrix tailored to the gain
state of the LAC at the time of the observation.
To guard against misinterpreting subtle spectral features caused by residual
uncertainties in the response matrix, we followed the  standard practice of
adding systematic errors of 1\% to the spectral (PHA) data.
We ignored PHA channels 1--2, which do not contain valid data, and
high energy channels where the sensitivity falls off, leaving an energy
range of 1.2--17.4 keV.  A few channels in some of the cluster spectra
were rebinned to improve statistics.

					\bigskip
					\centerline{\it
2.2\ \ Einstein SSS spectra		}
					\medskip

We extracted the SSS spectra of the four clusters from the HEASARC data base
(see Table 1 for the observation log).
Ice accumulated on the spectrometer entrance
window in a time variable manner (through a series of frost and
de-frost cycles), causing variable absorption of soft X-rays, so
each spectral data file requires its own response matrix. The latest
models of ice buildup were incorporated into the response  matrix generation
and
are advertised to work best for spectra in which the  ice parameter changes by
less than 10\% during the integration. The new ice models obviate the inclusion
of the previously recommended 2\% systematic errors (Christian $et$ $al.$
1992).
The statistical errors are far larger than any residual systematic errors. PHA
channels 1 and 86--94 were ignored, leaving an approximate energy range of
0.5--4.5 keV, which overlaps the low energy $Ginga$ LAC spectral coverage.
Some channels were rebinned to improve statistics.
To analyze the SSS spectra, we followed the detailed prescription given by
Drake, Arnaud and White (1992), which describes how to subtract the two SSS
background components (see Szymkowiak 1986 and WFJMA).

					\bigskip
					\centerline{
3.\ \ DATA ANALYSIS			}
					\medskip

The $Ginga$ LAC spectra of three of these clusters are among the eleven
analyzed previously by Hatsukade (1989), who fit a variety of spectral
models: thermal bremsstrahlung, bremsstrahlung plus one and
two emission lines (due to iron K$\alpha$ and K$\beta$), and collisional
ionization equilibrium (CIE) models based on the emissivity calculations of
Masai (1984). An additional power-law component was considered for two
clusters which are not in the current subset.

The $Einstein$ SSS spectra of these clusters were previously analyzed
by Mushotzky (1984), Mushotzky and Szymkowiak (1988) and WFJMA.
In re-analyzing nearly
all the clusters in the SSS database, WFJMA discovered the need for
considerable amounts of intrinsic absorption in most of the clusters with
cooling flows.  As mentioned above, this intrinsic absorption component tends
to be well in excess of the Galactic column, and was attributed to the
accumulation of cooling flow condensates over the lives of the cooling flows.

We used the XSPEC (version 8.33) software package to individually and
simultaneously analyze the SSS and LAC spectra for each cluster.
We fitted a variety of spectral models, including single- and dual-temperature
thermal models, and thermal plus cooling flow models, among others.  In models
with two emission components, we also allowed the abundances of the
two components to vary independently.  The results of the model fits are
listed in Tables 2-5, the entries of which are described in the notes to
Table 2; for brevity we will refer to specific model results by
table and line number: $e.g.$ T2.1 refers to line 1 of Table 2.
The results of these fits are now described for each cluster in turn.

					\vfil\eject        %SMALL
					\bigskip
					\centerline{\it
3.1\ \ Abell 496				}
					\medskip

Abell 496 is a Bautz-Morgan type I cluster with a central cD.  At a redshift
$z=0.0320$, the luminosity distance to A496 is 193 $h_{50}^{-1}$ Mpc and
$1^\prime=53\ h_{50}^{-1}$ kpc (where $h_{50}$ is the Hubble constant in units
of 50 km s$^{-1}$ Mpc$^{-1}$ and we take $q_0={1\over2}$,
so the angular diameter distance is 182 $h_{50}^{-1}$ Mpc).
The LAC field of view (FWHM)
encompasses $3.2\times6.4h_{50}^{-1}$ Mpc at this distance, while the
SSS field of view has a 318$h_{50}^{-1}$ kpc diameter.
The cluster appears dynamically relaxed, since it does not have
much velocity or spatial substructure (Zabludoff, Huchra \& Geller
1990, hereafter ZHG) and its IPC X-ray images have
smooth isophotes in the central regions.  The IPC images show no
serendipitous sources with sufficient flux to be significant
compared to that of the cluster.  The nominal center of the SSS field of view
is
0.24$^\prime$ from the IPC X-ray peak, while the nominal center of the
larger LAC field of view is 4.6$^\prime$ away from the IPC X-ray peak.

The presence of a cooling flow in A496 was first posited by Heckman
(1981) on the basis of its central H$\alpha$ emission, thought to be
due to cooling flow condensates cooling through 10$^4$ K. More recent
optical observations show extended emission (Cowie $et$ $al.$ 1983,
hereafter CHJY).
Nulsen $et$ $al.$ (1982) found a soft X-ray component in the
$Einstein$ SSS spectra for this cluster and estimated
a cooling accretion rate of $\sim200$ $\msunyr$.
Subsequent analysis of SSS spectra by Mushotzky (1984) and Mushotzky and
Szymkowiak (1988) indicated an accretion rate of 200--400 $\msunyr$, while
Canizares $et$ $al.$ (1988) derived upper limits of 200 and 310 $\msunyr$
from individual X-ray lines in FPCS spectra.
Further imaging analyses of IPC data by Arnaud and Fabian (1987) and
Thomas $et$ $al.$ (1987) derived a total accretion rate of $\sim$100--120
$\msunyr$
and a cooling radius of $\sim3.0^\prime$--$3.4^\prime$.  The larger of these
cooling radius estimates extends 0.4$^\prime$ beyond the SSS field of view.
However, in both of these imaging analyses, the age of the cooling flow was
assumed to be $2\times10^{10}$ yr.  Depending upon the cooling time criterion
adopted, this may lead to an overestimate of the cooling radius by as much as a
factor of $\gta2$ if the true age is $\lta10^{10}$ yr, given the observed
density profile.  Thus, the bulk of the cooling flow
emission may be within the SSS field of view, but we test this below.

					\medskip
					\centerline{\it
3.1.1\ \ Separate LAC and SSS Spectral Fits	}
					\medskip

We fitted a Raymond-Smith (hereafter denoted RS) isothermal model
for the X-ray emission from an optically-thin hot plasma in collisional
ionization equilibrium (Raymond \& Smith 1977)
to the LAC data for comparison with the results of Hatsukade (1989).
In XSPEC, the cosmic abundance of iron (by number) relative to hydrogen is
$-4.5$ dex, while Hatsukade (1989) adopted a value of $-4.4$ dex (Allen 1976).
Our abundance determinations (which are driven mostly by iron) will then
be $\sim$1.26 times larger than Hatsukade's for the same iron fraction.
For this comparison we adopted the same Galactic line-of-sight hydrogen
column density as Hatsukade (1989): $N_H=5\times10^{20}$ cm$^{-2}$
($N_H$ is used to parameterize the soft X-ray absorption due to
cool intervening Galactic material --- see Morrison and McCammon 1983).
Although the best fit is not good (the reduced $\chi^2$ per degree of freedom
$\chi^2_\nu=2.21$ for $\nu=19$ degrees of freedom; see line 1 of Table 2,
denoted T2.1),
we found the temperature and abundance to be well-determined,
with $kT=3.97$ (3.91--4.04) keV and $A=0.48$ (0.43--0.53), where
$A$ is the heavy element abundance relative to the cosmic value
($A$ is assumed to be the same for all elements)
and the values in parentheses are 90\% confidence intervals (see Table 2,
where the 90\% confidence limits vertically bracket the best-fit values).
Using Masai emissivities, Hatsukade (1989) found a slightly smaller
temperature, $kT=3.91$ (3.84--3.97) keV, and a slightly larger
iron abundance, $A=0.53$ (0.48--0.58) (adjusting upward a factor of 1.26),
which are nonetheless within the 90\% confidence intervals of the present
values.  If the column density of hydrogen is allowed to vary freely, it is
not well-determined.  Zero column provides the smallest $\chi^2$, reducing
$\chi^2$ by 10.6 while reducing the number of degrees of freedom $\nu$ by one.
That the previous model (see T2.1) has larger $\chi^2$ is due to the fact
that there is excess emission compared to the model at the lowest
energies, which becomes even more obvious when the minimum expected
absorption (due to the Galaxy) is taken into account in the model. This excess
emission is likely due to a relatively cool component analyzed further below.

We also analyzed the two SSS spectra (of three total) for which the ice varied
by less than 10\% in the course of the integration (see Table 1).
We first fitted isothermal models and found an excellent fit ($\chi^2_\nu=0.80$
for $\nu=148$; see T2.2) with a somewhat lower temperature,
$kT=3.68$ (3.08--4.34) keV, than the LAC data indicated,
although the LAC temperature is within the 90\% confidence upper limit.
The derived fractional abundance $A=1.50$ (0.94--2.28)  is supersolar,
substantially higher than that of the LAC spectrum.  Fixing the
abundance of the SSS spectra to match the half-solar value found in the
LAC spectrum produced a significantly worse fit, with $\Delta\chi^2=+15$ while
$\Delta\nu=+1$; an $F$-test shows that the previous
variable abundance fit is preferred with $>99.9$\% confidence.
This is a strong indication of enhanced abundances in the
SSS spectra relative to the more global LAC spectrum.
WFJMA also noted a roughly cosmic abundance for the cooling flow component in
the SSS data of this cluster, although no errors were given.  They fitted
a two-component emission model to the data, with one component being an
isothermal model at the $EXOSAT$-derived temperature of 4.8 keV
(Edge $et$ $al.$ 1990) and the other an isobaric cooling flow model.
We also found the large column density of X-ray absorbing material previously
found in the SSS spectra by WFJMA.  The best-fit value for this model is
$N_H=1.2$(1.0--1.5)$\times10^{21}$ cm$^{-2}$, which is twice the
Galactic value interpolated from Stark $et$ $al.$ (1992),
but we will show that the inferred amount is rather model dependent.

We also considered whether the high abundance seen in the SSS spectra may
be an artifact of systematic errors in the ice model adopted for the SSS
spectra.  The generation of SSS response matrices depends on only one variable,
the ice parameter appropriate for a given observation (see Table 1).
The ice parameter has a maximum uncertainty
of $\pm0.4$ during the time these data were taken (Christian $et$ $al.$ 1992).
We re-analyzed the SSS spectra using a response matrix characterized by
an ice parameter 0.4 larger than the best-estimate value used above.
This increases the amount of ice assumed to be on the SSS entrance window, so
the inferred internal absorption will be reduced:  in refitting our isothermal
model, the best-fit absorption was brought down to
$N_H=6.9$(4.7--9.4)$\times10^{20}$ cm$^{-2}$, which is
within $\sim50\%$ of the Galactic value.  The best-fit value of the
fractional abundance was reduced to $A=0.99$ (0.67--1.39), the 90\% confidence
range of which is still outside that of the more global LAC spectrum.
Although systematic errors may make the absolute value of the central
abundance somewhat insecure, the $existence$ of a central enhancement seems
more robust.

					\vfil\eject       %SMALL
					\medskip
					\centerline{\it
3.1.2\ \ Joint Spectral Fits		}
					\smallskip
					\centerline{\it
3.1.2.1\ \ Raymond-Smith Models		}
					\medskip

We next analyzed the LAC and SSS spectra jointly.
The simplest plausible model is isothermal and assumes the SSS and LAC spectra
are characterized by the same abundances and column density of cool absorbing
matter.  The normalizations
were allowed to vary independently for the two instruments, to allow
for their different fields of view.  This model fit poorly, with
$\chi^2=203.3$ for $\nu=169$ ($\chi^2_\nu=1.20$; see T2.3).
Allowing an additional spectral component with the same temperature and
abundance
but with an intrinsic absorbing column (at the redshift of the cluster)
in addition to a fixed Galactic column provided only a marginally improved fit.
(Note that for this and subsequent analysis of the other clusters, when we
are not directly
comparing to the work of Hatsukade 1989, we use Galactic values of $N_H$
derived from Stark $et$ $al.$ 1992, using K. Arnaud's {\smc getnh} program.)
A much better fit was provided by a two-temperature model in which the cooler
component was allowed an intrinsic absorbing column in addition to a fixed
Galactic column:  $\chi^2=163.3$ for $\nu=167$ (so $\chi^2_\nu=0.98$ and
$\Delta\chi^2=-40$ for $\Delta\nu=-2$; see T2.4).
In this model the cool component is attributed to the cooling flow
and its normalization was constrained to be the same in the SSS and LAC
fields of view, which assumes that the cooling flow emission is fully contained
in each.

Finally, we allowed the abundance to vary independently in each of
the two temperature components, which provided a significantly better fit,
with $\chi^2=153.6$ for $\nu=166$ ($\chi^2_\nu=0.93$; see T2.5).
The improvement over the single-abundance model of T2.4 ($\Delta\chi^2=-9.7$
for $\Delta\nu=-1$)
is significant with $>99.8\%$ confidence by the $F$-test.
The best-fitting such two-temperature model is shown with the data in Figure 1,
along with the residuals of the fit. This model has a hot component with
$kT=4.30$ (4.19--4.41) keV and a cooler component with $kT=1.64$
(1.39--2.01) keV. The temperature $\chi^2$ distributions are shown in Figure 2a
where the dashed horizontal lines represent the single-parameter
90\% and 99\% confidence
levels (corresponding to $\Delta\chi^2=+2.71$ and +6.63, respectively).
The temperature of the hotter component is
$\sim8$\% greater than that derived from the LAC spectrum alone.
The fractional abundance associated with the hotter component is $A_{hot}=0.47$
(0.42--0.53), while that of the cool component exceeds the cosmic value:
$A_{cool}=2.63$ (1.01--$>$5).  The $\chi^2$ distributions for the
abundances of the two temperature components are shown in Figure 3a, which
shows that the abundance of the cool component, while having large errors,
is nonetheless substantially greater than that of the hot component,
at a confidence level of $>$99\%.
The ratio of model fluxes in the SSS and LAC fields of view is consistent
with the ratio derived from IPC imaging data; the model fluxes were
calculated in the 0.16--3.5 keV energy band appropriate to the
background-subtracted, vignetting-corrected IPC images on the SAO CD-ROMs.
The cooler spectral component of this model has
an excess absorbing column density of 4.0(2.5--5.6)$\times10^{21}$ cm$^{-2}$,
nine times higher than the Galactic column.
The $N_H$ $\chi^2$ distribution is illustrated in Figure 4a, where the
Galactic value in the line of sight to A496 is also indicated.
If the ``intrinsic" absorption is distributed uniformly over the size of
the cooling flow (with radius $R_{cool}\approx100$ kpc),
there would be $\sim 9\times10^{11}\msun$ of cool material.
We considered a model which allowed the covering fraction of absorbing
material to vary and the subsequent best-fit value  was unity (so the fit
was not improved), with a 90\% confidence lower limit of $>83\%$ coverage.
However, since the intrinsic absorption is associated with the cool component
in this spectral model, its total angular covering fraction may be
substantially
less than this if the cool emission component itself is clumped and the
absorption is spatially correlated with the cool emission.

The column density associated with the cooler component of the
two-temperature fit is three times larger than that inferred from
the previous analysis of SSS spectra alone.
This is because the best-fit temperature in the previous isothermal
SSS analysis is larger than the temperature of the cooler component in
the two-temperature fit and $N_H$ is anticorrelated with the temperature:
for a fixed spectral shape at low energies, the effect of
reducing the temperature can be offset by increasing the absorbing column.
This anticorrelation is illustrated in Figure 5, which
shows the 1$\sigma$, 90\% and 99\% confidence contours for
$N_H$ {\it vs.} $T$ in the previous isothermal fit to SSS data alone.

To test whether the intrinsic absorption is largely confined to the
cooling flow (that is, within the SSS field of view), we considered a
two-temperature, two-abundance  model which allowed only one global
value of absorbing column to be fit.
The best-fitting such model is only marginally worse than the best model
illustrated above, having $\Delta\chi^2=+3.9$ for $\Delta\nu=0$ (see T2.6),
but the best-fit value of $N_H$ is more than twice the
Galactic value.  This value is very close to that found in the previously
described fit to the SSS data alone, which illustrates the insensitivity
of the LAC data to $N_H$.  Thus, we cannot demonstrate that the extra
absorption is confined to the SSS field of view. However, it is unreasonable
for the extra absorption to be due to an error in the estimate of the Galactic
component, since typical errors in interpolating Galactic values
on small angular scales from broad-beam surveys such as that of
Stark $et$ $al.$ (1992) are only of order $10^{20}$
cm$^{-2}$ (Elvis $et$ $al.$ 1986).

Although the model illustrated in Figures 1 and 2a--4a provide an excellent
fit to the LAC and SSS data,
inspection of the $\chi^2$ residuals for the best-fitting
two-temperature model above shows structure in the lowest-energy LAC
channels which may suggest that the LAC sees somewhat more cool
emission than the SSS.
To test whether the bulk of the cool component is confined within the SSS
field of view, we allowed the LAC and SSS normalizations for the cool
component to vary independently. The  resulting respective normalizations were
the same within their errors and the fit was not significantly
improved.  In an additional test,
we added a third temperature component to the model which could
contribute only to the LAC spectrum.  This LAC-only component ended
up being very cool ($kT\approx1$ keV) and significantly improved the fit,
reducing $\chi^2$ by $\sim12$. However, this cool component contributed only
1--2\% of the flux in the LAC passband, which is at the level expected from
calibration uncertainties and unresolved background sources; thus
there is little evidence of cool emission outside the SSS field of view.
$ROSAT$ imaging data will be incorporated in a subsequent paper.

					\medskip
					\centerline{\it
3.1.2.2\ \ Cooling Flow Models		}
					\medskip

In the previous two-temperature analysis we attributed the cooler of two
temperature components to the cooling flow at the center of A496.
However, a cooling flow spectrum is only roughly approximated by a
single-temperature RS thermal model.  We therefore investigated
models incorporating a more realistic cooling flow spectral component
in order to constrain the accretion rate and to assess the model-dependence
of the high metal abundances and absorption columns we deduce for the
cool component.  We adopted the cooling flow spectral model of Mushotzky
and Szymkowiak (1988), in its most recent implementation in XSPEC (v.8.33).
There are six variables in the cooling flow model:
the temperature from which gas cools $T_{high}$, the temperature to which
gas cools $T_{low}$, the accretion rate $\mdot$ (which is the normalization),
the fractional metal abundance $A$, the redshift $z$, and a slope
parameter $s$ for a temperature-dependent weighting of
the isobaric emission measure {\it vs.} temperature relation (the slope
is zero for an isobaric cooling flow).  The bulk of the cooling flow emission
will be radiated under roughly isobaric conditions in the outer half of the
cooling flow.  We find from XSPEC simulations that
pure cooling flow spectra (with $T_{low}$
below X-ray-emitting temperatures) are fit reasonably
well by isothermal RS models with temperatures $\sim2$--$3$
times smaller than $T_{high}$ (also see White and Sarazin 1987).  The cooler
component in the previously discussed two-temperature model thus
has roughly the temperature one expects from a cooling flow
starting at the higher of the two temperatures.

In our fits incorporating a cooling flow spectrum, we couple the
cooling flow model to a RS isothermal model for the cluster
emission exterior to the cooling flow.  As with the previous two-temperature
fits, we included a (fixed) Galactic absorption component and we associated
an additional (variable) ``intrinsic" absorption component with the cooling
flow.
In the cooling flow model we set $T_{high}$ equal to the temperature
of the exterior isothermal component, we fixed $T_{low}$ at 0.1 keV (below
the bulk of X-ray emitting energies), fixed the redshift,
and set the cooling flow slope parameter to zero (for isobaric flow).
The accretion rate, the abundances of the cooling flow and the exterior RS
components, and the intrinsic absorption freely varied, while $T_{high}$
was tied to and varied with the temperature of the exterior RS component.

We found an excellent fit for a model in which the cooling flow normalizations
were the same for the SSS and LAC: $\chi^2_\nu=0.91$ for $\nu=167$ (see
T2.7), which has $\Delta\chi^2=-1.9$ (for $\Delta\nu=+1$) compared
to the best two-temperature RS model with tied cool component normalizations.
The accretion rate was found to be 154 (127-188) $\msunyr$, consistent with
previous determinations, but with smaller errors.
The abundance of the cooling flow, $A_{cf}=3.79$ (1.58-$>$9), was found to
be significantly greater than that of the exterior gas,
$A_{hot}=0.25$ (0.16-0.37).  The $\chi^2$ distributions for the abundances
of the two spectral components are shown in Figure 6a.  Despite its large
error, the cooling flow abundance is greater than that of exterior gas at
$>99$\% confidence, which is consistent with the results of the previous
two-temperature analysis.
The ratio of model fluxes in the SSS and LAC fields of view are again
consistent
with that derived from IPC imaging data.
The best-fitting value of intrinsic $N_H$ is very
close to that of the previous analysis: $N_H=3.8$(3.2--4.6)$\times10^{21}$
cm$^{-2}$.   Using a similar RS plus cooling flow spectral model
(but with the cluster temperature fixed), WFJMA found a lower value for
the intrinsic column density: $N_H=2\times10^{21}$ cm$^{-2}$.
This disparity is largely due to WFJMA adopting the $EXOSAT$-derived
temperature of 4.8 kev (Edge $et$ $al.$ 1990), which is $\sim$15--20$\%$
higher than that derived by the higher quality $Ginga$ spectrum.
We further assessed the significance of the relatively enhanced cooling
flow abundance by considering a model with only a single variable abundance
for both the cooling flow and RS components.  This model (see T2.8) produced a
significantly larger $\chi^2$ ($\Delta\chi^2=+17$ for $\Delta\nu=+1$) than
the dual-abundance model of T2.7, which is preferred with $99.99\%$
confidence with the $F$-test.
To test whether the extra absorption component is confined to the SSS field
of view, we considered
a model with one variable global value of $N_H$ (see T2.9), rather than a
variable absorption component intrinsic to the cooling flow; this provided a
much worse fit ($\Delta\chi^2=+40$ for $\Delta\nu=0$) than the model of T2.7.
This suggests that the extra absorption is mostly confined to the SSS field of
view (but recall that a similar global-$N_H$ model employed in a previous
dual-RS fit [T2.6] did $not$ show such a pronounced degradation in the
quality of the fit).
A model allowing a partial covering fraction for the intrinsic absorption
produced a best-fit value of unity and a $90\%$ confidence lower limit
of $>85\%$ coverage.
To assess whether the cooling flow emission extends outside the SSS field
of view, we allowed the LAC and SSS cooling flow normalizations to vary
independently.  The fit was not significantly improved.

We conclude that A496 has a cool spectral component attributable
to a central cooling flow.  This cool component is associated
with a large column density of X-ray absorbing material, but these data
do not unambiguously show that it is confined to the SSS field of view.
There appears to be an abundance gradient in A496, with
abundances substantially larger at the center and declining outward.
The amount of metals in the cooling flow is far larger than what could
be produced by present levels of stellar mass loss, however, the metals
could have been produced during the integrated history of mass loss in
the central dominant galaxy and other galaxies in the core.
There does not appear to be much cool emission outside the SSS field of view.

%==============================================================================
					\bigskip
					\centerline{\it
3.2\ \ Abell 1795			}
					\medskip

Abell 1795 is a Bautz-Morgan type I cluster with a central cD.  At a redshift
$z=0.0620$, the cluster has a luminosity distance of 377 $h_{50}^{-1}$ Mpc,
and 1$^\prime=96$ $h_{50}^{-1}$ kpc (for an angular diameter distance of 335
$h_{50}^{-1}$ Mpc).
The central cD has a significant peculiar velocity, 370 km s$^{-1}$,
given its cluster velocity dispersion of $\sigma=773$ km s$^{-1}$ (ZHG,
Gebhardt and Beers 1991); more general velocity substructure is not
apparent in its velocity histogram (ZHG).  In the inner parts of the
central cD, McNamara \& O'Connell (1992) found a pronounced optical
color gradient relative to that expected from a normal
giant elliptical and attributed this to ongoing
star formation in the surrounding cooling flow.

$Einstein$ IPC X-ray imaging analyses find accretion rates ranging from
$\sim$340--400 $\msunyr$ (Stewart $et$ $al.$ 1984, Arnaud and Fabian 1987).
This range is similar to that deduced from
SSS spectra by Mushotzky (1984) and Mushotzky and Szymkowiak (1988):
425--560 $\msunyr$.
Arnaud and Fabian (1987) find a cooling radius of $\sim2.2^\prime$,
well within the SSS field of view.  The long optical filament discovered
by CHJY (extending to $\sim45^{\prime\prime}$ from near the center),
which may be a cooling flow condensate,
is also well within the SSS field of view.

IPC images of A1795 show that there are several compact sources
in the LAC field of view (but outside the SSS field of view).
The brightest of these sources is a Seyfert 1 galaxy (EXO 1346.2+2645).
The flux from these sources is $\lta 3\%$ of the total in the $Einstein$
passband, so their presence does not effect these spectral fits significantly.
The nominal center of the SSS pointing is within 0.6$^\prime$ of the IPC X-ray
peak, while the nominal center of the LAC pointing is within 8$^\prime$ of
the X-ray peak.
					\vfil\eject      %SMALL
					\medskip
					\centerline{\it
3.2.1\ \ Separate LAC and SSS Spectral Fits	}
					\medskip

The previous LAC analysis of Hatsukade (1989) found a collisional ionization
equilibrium model temperature of $kT=5.34$ (5.23--5.45) keV and an (adjusted)
fractional iron abundance $A=0.46$ (0.41--0.51),
for an assumed Galactic hydrogen column of $N_H=1.7\times10^{20}$ cm$^{-2}$.
We fitted a RS isothermal model to
the LAC spectrum alone with the same value of $N_H$ and found a
slightly higher temperature and lower
abundance: $kT=5.57$ (5.45--5.69) keV and $A=0.43$ (0.38--0.48),
with 90\% confidence intervals which overlap those of Hatsukade
(see T3.1).
If the hydrogen column density is allowed to vary freely,
it is not well-determined; the smallest $\chi^2$ is associated
with zero column, but $\chi^2$ is reduced by only 2.4 for
$\Delta\nu=-1$.

Of the six SSS spectra available for A1795, we analyzed three.  Two of the
omitted SSS spectra had ice variations of $\sim50\%$ in the course of their
integrations and the third omitted spectrum had twice as much ice as the
remaining three (see Table 1).  Our isothermal model fits to the SSS spectra
have significantly lower temperatures than found in the LAC spectrum.  With the
abundance free to vary, the best-fitting temperature of an adequate fit
($\chi^2_\nu=1.07$ for $\nu=225$) is 3.67 (3.26--4.14) keV.
The abundance is very low and poorly constrained, with a 90\% confidence
upper bound of 0.24 cosmic and no lower bound (see T3.2).
However, fixing the abundance to that indicated in the LAC spectrum
significantly worsened the fit ($\Delta\chi^2=+9$ for $\Delta\nu=+1$;
$\chi^2_\nu=1.11$); an $F$-test shows that the previous fit with
variable abundance is significantly better at the $99.5\%$ confidence level.

					\medskip
					\centerline{\it
3.2.2\ \ Joint Spectral Fits		}
					\smallskip
					\centerline{\it
3.2.2.1\ \ Raymond-Smith Models		}
					\medskip

Having the SSS spectra characterized by a temperature lower than that
of the LAC spectrum motivated a two-temperature model jointly
fit to the SSS and LAC data.  This provided an improvement in $\chi^2_\nu$.
The model is nearly the same type as used in A496, incorporating a fixed
Galactic absorption column and an additional variable absorption component
associated with the cooler of the two temperature components.  As before,
the additional ``intrinsic" absorption was assumed to
be at the redshift of the cluster. In this
cluster, however, the abundance of the cool component was tied to that
of the hot component since the SSS data provided such a poor constraint on
the abundance.  In modeling a cool component associated with the central
cooling flow, we can safely assume that the SSS field of view
encompasses all of the cooling flow emission, given the distance to A1795.
(There is still the possibility of cool emission in the outer parts of the
cluster where it could be detected by the LAC, but only a small projected
portion of which would be detected by the SSS.)
The best-fitting model is quite good ($\chi^2_\nu=0.99$, $\nu=244$) and
is illustrated with the data in Figure 7. The temperatures of the two
components are $kT=5.73$ (5.59--5.87) keV and $kT=0.74$ (0.58--0.90) keV
(see T3.4); their $\chi^2$ distributions are shown in Figure 2b.
The overall fractional abundance is $A=0.43$ (0.38--0.48) and its $\chi^2$
distribution is illustrated in Figure 3b. When we allowed the abundance of the
cool component to vary independently of that of the hot component, we
found that the fit was not improved and the cool abundance was poorly
constrained.  Thus, these RS models provided no
evidence of centrally enhanced abundances.
The ratio of model fluxes in the SSS and LAC fields of view is consistent with
that derived from IPC imaging data.
A substantial absorbing column
is associated with the cool component, $N_H=5.1$(2.4-7.6)$\times10^{21}$
cm$^{-2}$, which is $\sim45$ times larger than the line-of-sight Galactic
column; Figure 4b shows the $N_H$ $\chi^2$ distribution.  As we show below,
this high value of $N_H$ is very model-dependent, with much lower values
provided by models incorporating cooling flow spectra.

This model provided a significantly better fit than simpler alternatives, the
simplest being characterized by a single temperature, abundance and absorbing
column.  The best-fitting such model (see T3.3)
had $\Delta\chi^2=+42$ (for $\Delta\nu=+2$) relative to the previous
two-temperature model.
We further tested the significance of the cooler component by forcing the
temperatures of the two components to be the same, but allowing one of
the isothermal components to have intrinsic absorption at the cluster
redshift (in addition to an overall absorption fixed at the Galactic value).
The best-fitting such model had $\Delta\chi^2=+46$ (for $\Delta\nu=+1$)
compared to the previous two-temperature model with intrinsic absorption.
The need for a cooler component is evidently extremely significant:
the $F$-test shows that the improvement provided by the second temperature
component is significant with $>99.99\%$ confidence.
We tested whether the ``intrinsic" absorbing column is confined
to the SSS field of view by forcing the two-temperature model to have only
one global value of $N_H$.  The best-fitting such model had
$N_H=3\times10^{20}$ cm$^{-2}$, three times the Galactic value,
but provided a somewhat poorer fit, with $\Delta\chi^2=+6.2$ for $\Delta\nu=0$.
The poorer fit is weak evidence for the extra absorption being confined
to the SSS field of view, but the
value of $N_H$ is consistent with that indicated in the isothermal
analysis of the SSS data alone (see T3.2), so the LAC data again provide a
poor constraint on $N_H$ and its coverage.
We also considered a partial coverage model for the intrinsic absorption
component (in a spectral model otherwise like that of T3.4)
and found the best-fit value of the covering fraction to be
unity (so the fit was not improved),
with a $90\%$ confidence lower limit of $>80\%$ coverage.

As for A496, we tested whether the LAC saw significantly more cool emission
than was in the SSS field of view by adding a third
RS component to the LAC spectral model.  The resulting best-fit
$\chi^2$ was not significantly improved and the contribution of the (cool)
third component to the LAC passband was only $\sim1\%$.
As we saw before in A496, this is at the level expected from calibration
uncertainties and unresolved background sources.

					\medskip
					\centerline{\it
3.2.2.2\ \ Cooling Flow Models		}
					\medskip

Since a central cool component was strongly indicated above,
we also considered a spectral model incorporating a cooling flow component
in order to constrain its accretion rate and assess the model dependence
of the intrinsic $N_H$.  Given the lack of evidence of enhanced abundances
in the SSS spectra relative to the LAC spectrum, we initially tied the
abundances of
the two spectral components together.  The cooling flow normalization $\mdot$
was taken to be the same in the LAC and SSS spectra.
The best-fitting model is quite good, although not quite as good as the
best-fitting two-temperature model described above: in this case,
$\chi^2/\nu=252.6/245$ ($\chi^2_\nu=1.03$, see T3.5), so compared to the better
fitting two-temperature model (T3.4), $\Delta\chi^2=+10$ for $\Delta\nu=+1$.
The exterior temperature, $kT=6.07$ (5.85--6.33) keV, is $\sim9\%$
larger than in the previous two-temperature RS
model. The fractional abundance $A=0.45$ (0.40--0.50)
is the same as derived previously.
The intrinsic absorption, $N_H=8.9$(5.8--11.5)$\times10^{20}$ cm$^{-2}$ is
$\sim6$ times smaller than that associated with the cool component
of the previous model; this is still $\sim8$ times larger than the Galactic
column in this line of sight and $\sim10\%$ larger than the value found by
WFJMA.
The accretion rate is consistent with previous estimates:
$\mdot=411$ (290--533) $\msunyr$.

When we allowed the abundances of the cooling flow and RS
components to vary independently, a slightly better fit was achieved, with
$\Delta\chi^2=-3.9$ for $\Delta\nu=-1$ (see T3.6); this improvement is
significant at the $95\%$ level by the $F$-test.
The cooling flow abundance
$A_{cf}=0.58$ (0.46--0.71) ended up significantly larger than that of the hot
component $A_{hot}=0.14$ (0.04--0.37), with non-overlapping $90\%$ limits,
but $A_{hot}$ is poorly determined. The $\chi^2$ distributions for the
abundances of the two spectral components are shown in Figure 6b.
The ratio of model fluxes in the SSS and LAC fields of view is consistent with
that derived from IPC imaging.
We also tested whether this model allows the extra absorbing column to be
confined to the SSS field of view: allowing only a global
value of $N_H$ led to a slightly worse fit, with $\Delta\chi^2=+1.6$ for
$\Delta\nu=0$.  The best-fit global value of $N_H$ is $\sim4$ times
higher than the Galactic value (see T3.7) and several times higher than
the plausible uncertainty in the Galactic value.  However, given the comparable
quality of the fit, we cannot prove that the extra absorption is confined
to the SSS field of view.
We also considered a partial covering model for the intrinsic absorption
(in a model otherwise like that of T3.6)
and found the best-fit value to be unity (so the fit was not improved),
and the $90\%$ confidence lower-limit was $>45\%$ coverage.

We conclude that there is a cool spectral component attributable to a
cooling flow in A1795.  In contrast to A496, the evidence is mixed on
there being a significant central abundance enhancement:  the
dual-temperature RS models show no evidence of a significant enhancement,
while the poorer-fitting cooling flow models do show evidence of an
enhancement.
We found the amount of cool, X-ray absorbing material first found by WFJMA
to be rather model-dependent, but still very substantial.
We found no evidence in this data of the LAC
seeing significantly more cool emission than the SSS.

					\bigskip
					\centerline{\it
3.3\ \ Abell 2142			}
					\medskip

Abell 2142 is a Bautz-Morgan type II cluster, with a dominant
pair of giant ellipticals at its center (Rood-Sastry type B),
one of which has multiple nuclei (Hoessel 1980).  With a
redshift $z=0.0899$, the cluster is the most distant in our
sample: it has a luminosity distance of
551 $h_{50}^{-1}$ Mpc, so 1$^\prime=132$ $h_{50}^{-1}$ kpc (for an angular
diameter distance of 464 $h_{50}^{-1}$ Mpc).

X-ray imaging analysis indicated the presence of a cooling flow with
an accretion rate of $\sim55$ $\msunyr$ (Arnaud and Fabian 1987) and
a cooling radius of $\sim0.7^\prime$.  Previous analysis of SSS spectra
provided an upper limit to the cooling flow accretion rate of
$\mdot < 540$ $\msunyr$ (Mushotzky and Szymkowiak 1988).  Since the cooling
flow is contained in only $\sim5$\% of the SSS field of view, its emission
is heavily diluted by hotter exterior gas.

The IPC image has three other sources in the $Ginga$ field of view
which are outside the SSS field of view:
a K5 star (SAO 84114) and a Seyfert 1 galaxy (1556+259) very close together
on the sky (Reichert $et$ $al.$ 1982) and a QSO (1557+272).
They contribute only $\sim2\%$ of the flux in the IPC field, so their
contribution is negligible compared to that of the cooling flow.
The nominal center of the SSS pointing is 0.53$^\prime$ from the IPC
X-ray peak, while that of the LAC is 3.6$^\prime$ away.

					\vfil\eject      %SMALL
					\medskip
					\centerline{\it
3.3.1\ \ Separate LAC and SSS Spectral Fits	}
					\medskip

Hatsukade (1989) previously found a temperature of $kT=8.68$ (8.48--8.88) keV
and an abundance of 0.35 (0.30--0.40) cosmic in the LAC spectrum.
Our best-fitting RS isothermal model using the same value of $N_H$
has a slightly larger temperature $kT=8.99$ (8.78--9.18) and virtually
identical fractional abundance $A=0.34$ (0.29--0.39) (see T4.1).
This isothermal model provides a formally excellent fit, with $\chi^2_\nu=0.85$
for $\nu=21$ (see T4.1).  When $N_H$ is allowed to vary freely,
it is poorly determined and $\chi^2$ is not significantly improved.

We also analyzed the three SSS spectra available for this cluster. We first
fitted isothermal models and found an excellent fit ($\chi^2_\nu=0.94$,
$\nu=248$; see T4.2) with a temperature of $kT=7.81$ (6.74--9.01)
keV; the 90\% confidence upper limit encompasses the LAC-derived temperature.
However, the derived fractional abundance $A=1.97$ (1.08--3.26) is much
higher than that indicated in the LAC spectrum and the hydrogen column of
8.4(7.2--9.5)$\times10^{20}$ cm$^{-2}$ is about twice the Galactic value.
A $\chi^2$ contour of $kT$--$A$ shows that the abundance enhancement is
not the result of temperature uncertainties, since there is a firm 99\%
confidence lower limit of $A\gta0.5$.
Fixing the model abundance to match the smaller value found in
the LAC spectrum produced a significantly worse fit ($\Delta\chi^2$=+11.6,
for $\Delta\nu=+1$), which is rejected relative to the former model (in T4.2)
at the $99.95\%$ confidence level by the $F$-test.

As for A496, we assessed whether the super-solar abundance may be due
to systematic errors in the SSS ice model.  The maximum uncertainty in the ice
parameter for these data is again $\pm0.4$, so we again employed
response matrices which were characterized by ice parameters 0.4 larger
than the best-estimate value used above.  The subsequent best-fitting
isothermal
RS model had a smaller abundance, $A=0.71$ (0.44--1.04), however its
90\% confidence range is still outside that of the more global LAC
spectrum.  This is a strong indication that abundances are enhanced
in the SSS field of view relative to exterior gas.
The absorbing column was reduced to
$N_H=2.8$(1.5--4.1)$\times10^{20}$ cm$^{-2}$, which is consistent with
the Galactic value.

					\medskip
					\centerline{\it
3.3.2\ \ Joint Spectral Fits		}
					\smallskip
					\centerline{\it
3.3.2.1\ \ Raymond-Smith Models		}
					\medskip

In a joint analysis of the SSS and LAC spectra, we found a formally good fit
for the simplest plausible RS model, characterized by a single
temperature, abundance and column density ($\chi^2_\nu=0.98$, $\nu=271$;
see T4.3), although the best-fit ``global" $N_H$ is twice the Galactic column.
Since the SSS data alone seem to require roughly cosmic abundances, while the
LAC data require abundances which are one third as much, we also considered
a dual RS model with the same temperature but with independently
varying abundances.
We found a somewhat improved fit
($\Delta\chi^2=-6.6$ for $\Delta\nu=-2$;
see T4.4), but the abundance of the dominant RS
component (with normalization dictated by the LAC) vanished.
This means that virtually all of the line emission seen by the LAC could
conceivably come from within a region defined by the SSS field of view, leaving
few metals to the exterior.  The fractional abundance of the (cooler) component
within the SSS field of view is $A_{cool}=2.35$ (0.99--$>$4),
while the 90\% upper limit to the hot component fractional abundance is
$A_{hot}\lta0.20$.

A two-temperature model with separately variable abundances (see T4.5)
provided an
additional improvement on this already acceptable fit, giving $\chi^2_\nu=0.95$
or $\Delta\chi^2=-5.4$ for $\Delta\nu=-1$ relative to the previous model
in T4.4; this improvement is significant at the $98\%$ level by the $F$-test.
Once again, however, the
dominant component (now hotter) has a vanishing abundance (with a 90\%
confidence upper limit of $A_{hot}\lta0.22$ solar).
This model and the data are shown in Figure 8.
The $\chi^2$ distributions for the temperatures of the
two RS components are shown in Figure 2c.
The temperature of the hot component is $\sim10$\% greater than
that derived from the LAC data alone.  The $\chi^2$ distributions
for the hot and cool abundances are shown in Figure 3c,
where it is clear that the abundances differ with $>99\%$ confidence.
This model also required an intrinsic absorbing column of $\sim5$ times the
line-of-sight Galactic value: $N_H=2.0$(1.3--3.9)$\times10^{21}$ cm$^{-2}$;
the $N_H$ $\chi^2$ distribution is shown in Figure 4c.
The ratio of model fluxes in the SSS and LAC fields of view is consistent with
that derived from IPC imaging.

Fixing the abundance of the hot component to be equal to that derived
from the LAC data alone provided an acceptable fit ($\chi^2_\nu=0.97$
for $\nu=269$; see T4.6) although with some degradation
compared to the previous model with vanishing abundances: $\Delta\chi^2=+6.3$
for $\Delta\nu=+1$.  The fractional abundance of the cool component remained
relatively high, with a 90\% confidence interval which just touched
that of LAC data analyzed alone: $A_{cool}=0.97$ (0.39--2.35).
Finally, the need for two abundances was further tested by forcing the
cool and hot components to have the same abundances.  The best fit (see T4.7)
had $\Delta\chi^2=+8.9$ for $\Delta\nu=+1$ relative to the dual-abundance
model in T4.5; the need for the additional abundance component is
significant at the $99.7\%$ confidence level by the $F$-test.

We tested whether the substantial intrinsic absorbing column required above
is confined to the SSS field of view by fitting a two-temperature,
two-abundance
model with only one (variable) global column density.  The resulting best fit
is
nearly as good as that detailed in T4.5 ($\Delta\chi=+1.3$ for
$\Delta\nu=0$), so we cannot demonstrate that the absorption is confined
to the SSS field of view.
However, the best-fitting $N_H$ is twice as high as the Galactic
value, eight times larger than its likely uncertainty.
We also explored a partial covering fraction model for the intrinsic
absorption:
the best fit value of the covering fraction was unity
(so the fit was not improved),
with a $90\%$ confidence lower limit of $\sim60\%$.

					\medskip
					\centerline{\it
3.3.2.2\ \ Cooling Flow Models		}
					\medskip

We also investigated spectral models incorporating a cooling flow component
in order to constrain its accretion rate.
A RS model plus the standard cooling model employed for
the previous clusters provided about as good a fit as the best-fitting dual
RS models, with $\chi^2_\nu=0.95$ for $\nu=269$ (see T4.8).
The abundance of the RS model exterior to the cooling
flow was again very low, with $A_{hot}=0.12$ (0--0.32), while the cooling
flow component has a super-solar abundance with large errors:
$A_{cf}=3.6$ (0.93--$>$5.6); the $\chi^2$ distributions for the abundances
of these two spectral components are shown in Figure 6c.
The accretion rate is $\mdot=439$ (345--545) $\msunyr$, substantially higher
than that derived from the imaging analysis of Arnaud and Fabian (1987).
Substantial intrinsic absorption is again indicated, with a higher column
than in the previous dual RS fits:
$N_H=3.5(3.4$--$5.5)\times10^{21}$ cm$^{-2}$.  This is several times higher
than the intrinsic column found by WFJMA.  The SSS/LAC model flux ratio for
this model is consistent with that derived from IPC imaging.

Forcing the abundances of the cooling flow and exterior RS
components to be equal degraded the quality of the fit somewhat
($\Delta\chi^2=+6.4$ for $\Delta\nu=+1$; see T4.9); the former dual-abundance
model (T4.8) is preferable at the $99\%$ confidence level by the $F$-test.
As we did for the dual RS models, we also tested whether the
intrinsic absorbing column is confined to the SSS field of view:  still keeping
the abundances of the two spectral components tied, we allowed only one
globally varying absorbing column and found little additional
degradation in the best fit (see T4.10), so we cannot prove that
the absorption is confined to the SSS field of view.  Nonetheless, the
best-fitting global absorbing column was about twice the nominal Galactic
value, just as we found in the dual RS test, so the excess absorption is
several times larger than the likely error in the Galactic column.
The best-fitting accretion rate,
$\mdot=177$ $\msunyr$, is less than half that derived in models which allowed
for an intrinsic absorption component.  We also considered a partial covering
fraction model for the intrinsic absorption; in a model similar to that
of T4.8, we found a best-fit value of unity for the covering fraction
(so the fit was not improved), with a $90\%$ confidence lower limit of $>80\%$.

We conclude from these various spectral model fits that there is strong
spectral evidence for a relatively cool central component in A2142 and
that its abundance is likely to be enhanced relative to that of
exterior gas, although its absolute level is somewhat uncertain due to
SSS ice parameter uncertainties.
We found that the amount of intrinsic absorption previously found by
WFJMA, although large, is rather model-dependent.
The spectrally-derived cooling accretion rate depends
sensitively upon how the intrinsic absorption is distributed.

					\bigskip
					\centerline{\it
3.4\ \ Abell 2199			}
					\medskip

A2199 is a cD cluster of Bautz-Morgan type I.  The central cD, NGC 6166, has a
triple nucleus (Hoessel 1980), one of which contains unresolved H$\alpha$
emission (CHJY).  At a redshift $z=0.0309$, the luminosity distance is 187
$h_{50}^{-1}$ Mpc, the closest in our sample,
and $1^\prime=51$ $h_{50}^{-1}$ kpc (for an angular diameter distance of 176
$h_{50}^{-1}$ Mpc).  There is pronounced asymmetry in
the cluster's histogram of galaxy velocities (ZHG) and NGC 6166 has
a peculiar velocity of $\sim380$ km s$^{-1}$ relative to the cluster mean.
ZHG found this peculiar velocity to be significant compared to the cluster
velocity dispersion of 794 km s$^{-1}$, but Gebhardt and Beers (1991),
using a more robust statistical test, suggest otherwise.
McNamara and O'Connell (1992) found photometric
evidence of ongoing star formation in the central cD: a modest optical
color gradient relative to the color profiles of ``normal", non-star-forming
template ellipticals.

X-ray imaging studies indicate the presence of a central cooling flow
with an accretion rate  of
$\sim100$--$220$ $\msunyr$ (Stewart $et$ $al.$ 1984; Arnaud and Fabian 1987;
Thomas $et$ $al.$ 1987).  Cooling radius estimates range from
$3.2^\prime$--$6.7^\prime$ (Arnaud and Fabian 1987; Thomas $et$ $al.$ 1987),
extending beyond the $3^\prime$ radius of the SSS field of view.
However, as noted for A496, if the system age were $\lta10^{10}$ yr,
rather than the $2\times10^{10}$ yr adopted in these studies,
the cooling radius may be within the SSS field of view.
Mushotzky (1984) and Mushotzky and Szymkowiak (1988) deduce
an accretion rate of $\sim45$--$60$ $\msunyr$ from SSS spectra.
No serendipitous sources significant enough to affect this study
appear in the IPC image of this cluster.  The nominal center of the SSS
field is 0.10$^\prime$ from the IPC X-ray peak, while that of the LAC
is 10.7$^\prime$ away.

					\medskip
					\centerline{\it
3.4.1\ \ Separate LAC and SSS Spectral Fits	}
					\medskip

A RS isothermal spectral model fit to the LAC data alone provides an
adequate fit ($\chi^2_\nu=1.10$ for $\nu=22$) with a temperature $kT=4.48$
(4.41--4.54) keV and a fractional abundance $A=0.43$ (0.39--0.46) (see T5.1).
Fixing the absorbing column to the very low Galactic value
($\sim9\times10^{19}$ cm$^{-2}$), rather than allowing it to be fit freely,
increased $\chi^2$ by 1.4 (while $\Delta\nu=+1$), so there was little
significant effect.

There are three SSS spectra and all have good ice parameters.  The
best-fitting isothermal model provided a mediocre fit, with $\chi^2_\nu=1.15$
for $\nu=248$ (see T5.2).
The temperature $kT=3.34$ (3.05--3.70) keV is 25\% lower than that
derived from the LAC spectrum and the respective $90\%$ confidence $\chi^2$
limits do not overlap.  The fractional abundance, $A=0.61$ (0.44--0.82)
may be somewhat higher than that indicated in the LAC spectrum, but the
$90\%$ confidence intervals overlap slightly.  Intrinsic absorption
is strongly indicated in the SSS spectra, with
$N_H=7.9$(6.8--9.1)$\times10^{20}$ cm$^{-2}$,
which is $\sim9$ times larger than the Galactic column in this line of sight.

					\medskip
					\centerline{\it
3.4.2\ \ Joint Spectral Fits		}
					\smallskip
					\centerline{\it
3.4.2.1\ \ Raymond-Smith Models		}
					\medskip

In joint spectral fits, the simplest reasonable model employing RS
components is characterized by a single temperature, abundance and absorbing
column for both spectral data sets.  This model did not provide a good fit,
having $\chi^2_\nu=1.31$ (see T5.3).

Having the SSS spectra characterized by a cooler temperature than the
LAC spectrum motivated a joint, two-temperature fit to both data sets.
As before, the cooler component was assumed to be completely contained
within the SSS field of view, so it had the same normalization for
both the SSS and LAC spectra.  With the abundances of the two
temperature components tied together, the resulting best fit is shown in
Figure 9 and has $\chi^2_\nu=1.15$ for $\nu=271$ (see T5.4).
The higher temperature component, with $kT=5.28$ (4.86--5.46) keV, is
$\sim20\%$ hotter than indicated by the LAC spectrum alone.
The temperature of the cooler component, $kT=3.15$ (1.82--3.36) keV, is
consistent with the temperature found in the isothermal analysis
of the SSS spectra alone (see T5.2).
The $\chi^2$ distributions for the two temperature components are shown in
Figure 2d, while that of the (tied) abundance is in Figure 3d.
The intrinsic absorption has the same large column that was found in the
previous SSS analysis, $N_H=8.1$(7.1--24.7)$\times10^{20}$ cm$^{-2}$,
and its $\chi^2$ distribution is shown in Figure 4d.
The ratio of model fluxes in the SSS and LAC fields of view is consistent
with that derived from IPC imaging data.
When we allowed the fractional abundances of the two temperature components
to vary independently, the fit was not significantly improved and the
best-fit abundance of the cool component was only $\sim25\%$ higher than that
of the hot component --- their 90\% confidence limits also overlapped
substantially.  Isothermal models with either one or two abundances
provided much worse fits.  A variable covering fraction model for the
intrinsic absorption had a best-fit value of unity for the covering
fraction, with a $90\%$ confidence lower limit of $73\%$.

A better fit was provided by a two-temperature model with a single global
absorption component (see T5.5), rather than the combination
of fixed Galactic $N_H$ plus variable intrinsic absorption described above.
However, the SSS/LAC flux ratio in this model $exceeds$ that derived from
IPC imaging data (by $5\%$), so this model is not preferred: since the LAC
field
of view exceeds that of the IPC, the SSS/LAC model flux ratio
should be less than or equal to that derived from the IPC.

To assess whether the central component spreads significantly outside the
SSS field of view, we allowed the cool component normalization to vary
independently in the SSS and LAC spectral groups:  the best-fit values did
not differ significantly.  We also added a third RS component
which was allowed to be seen by the LAC alone, and this also failed to
significantly improve $\chi^2$.  The amount of flux this third
(cool) component ended up contributing to the LAC passband was $\sim1\%$.
This is at the level expected from uncertainties in calibration and
unresolved background sources.

					\medskip
					\centerline{\it
3.4.2.2\ \ Cooling Flow Models		}
					\medskip

As previously, we assessed the model-dependence of the abundance and intrinsic
$N_H$ estimates by considering a spectral model incorporating a cooling flow
component.
The cooling flow was again assumed to be isobaric, with an initial temperature
equal to that of the exterior isothermal atmosphere, and the cooling flow
normalization (accretion rate) was the same for the SSS and LAC spectra.
This cooling flow model provided a better fit than any of the previous
two-temperature models, having $\chi^2_\nu=1.05$ for $\nu=271$ (see T5.6).
The best-fitting exterior temperature, $kT=4.74$ (4.65--4.83)
keV, is $\sim6$\% larger than that derived from the LAC spectrum alone.
The abundances of the two spectral components were allowed to vary
independently and the best-fitting cooling flow abundance was near cosmic,
with $A_{cf}=1.15$ (0.65--2.11) while the exterior gas had
$A_{hot}=0.37$
(0.30--0.43), somewhat smaller than that derived from the two-temperature fits
(see Figure 6d for its $\chi^2$ distribution). The best-fitting intrinsic
absorption column, $N_H=3.0$(2.7--3.4)$\times10^{21}$ cm$^{-2}$, is $\sim4$
times larger than that associated with the previous two-temperature model.
This is also twice as high as the value found by WFJMA.
The accretion rate associated with the cooling flow is $\mdot=150$ (129--173)
$\msunyr$.  The ratio of model fluxes in the SSS and LAC fields
of view is consistent with that derived from IPC imaging data.
The quality of this fit suggests that the bulk of the cooling
flow emission is indeed within the SSS field of view.  When the cooling flow
normalizations were allowed to
vary independently in the LAC and SSS data sets (to test whether there
is significant cooling flow emission outside the SSS field of view) we found
the normalizations did not differ significantly and the $\chi^2$ was not
significantly improved.
To test the reality of the enhanced cooling flow abundances, we tied the
abundances of the two spectral components together.  The resulting best fit was
significantly worse, having $\Delta\chi^2=+7$ for $\Delta\nu=+1$ (for
$\nu=272$;
see T5.7).  The $F$-test shows that the need for the additional abundance
component in the previous model (T5.6) is significant at the 99\% level.
A model allowing a partial covering fraction
for the intrinsic absorption (but like the model in T5.6 in other respects)
produced a best-fit value of unity for the covering fraction, with a
$90\%$ lower limit of $>85\%$ coverage.

To test whether the intrinsic absorption component is truly intrinsic, we
fitted
a model with a single, variable, global absorption component.  The resulting
best-fit model (see T5.8) had $\Delta\chi^2=+10$ compared to the previous
best-fitting model and the absorption was still $\sim9$ times larger than the
Galactic value, which is unlikely to be wrong by such a large factor.
This poorer fit suggests that the intrinsic absorption is not globally
distributed, but mostly confined within the SSS field of view.

We conclude that there is strong evidence for a cool emission component with
intrinsic absorption in A2199.  Evidence for an abundance gradient is
model-dependent: the dual RS models did not have compelling evidence
for an abundance gradient, but the better-fitting models with cooling flow
spectra did have cooling flow abundances significantly enhanced compared
to exterior gas.

					\bigskip
					\centerline{
4.\ \ SUMMARY AND DISCUSSION			}
					\medskip

We jointly analyzed $Ginga$ LAC and $Einstein$ SSS spectra for four
galaxy clusters thought to contain cooling flows.  Each cluster's
spectra showed strong evidence for a relatively cool spectral component
associated with the central cooling flow.  The inclusion of these cool
components caused the temperatures derived for the more dominant hotter
components to be as much as $\sim20\%$ larger than those derived from
isothermal analyses, with $\sim10\%$ increases being more typical.
We found no compelling evidence for significant amounts of cool emission
outside the central regions encompassed by the SSS field of view ($i.e.$
within a 3$^\prime$ radius of the center).
We also explored the model-dependence in the amount of cool X-ray absorbing
matter in each of the clusters; this cool absorbing material
was first discovered by WFJMA to be ubiquitous in cooling
flow clusters.

All four clusters show evidence of centrally enhanced metal abundances, with
varying degrees of model dependence and statistical significance:
the evidence is statistically
strongest for A496 and A2142, weaker for A2199 and weakest for A1795.
In A496, both dual-temperature RS spectral models and models
incorporating cooling flow spectra provide the best fits when abundances
are enhanced within the SSS field of view (at $F$-test confidence levels of
$\ge99.85\%$ when compared to a single-abundance model).
Central abundance enhancement is also seen in both kinds
of spectral models for A2142 (at $F$-test confidence levels of $\ge99\%$
when compared to a single abundance model).
In A2199, dual-temperature RS models see no
significant central abundance enhancement, while the better-fitting
cooling flow models do produce central abundance enhancements (the latter at
an $F$-test significance level of $99\%$ compared to a single-abundance model).
Dual-temperature RS models for A1795 also showed no compelling evidence for a
significant abundance gradient, but these models provided better overall fits
than did models incorporating cooling flows, which did show evidence of a
central abundance gradient (at an $F$-test confidence level of $95\%$ relative
to a single abundance model).

Previous observations of abundance distributions in clusters have also
produced rather disparate results.
In M87/Virgo, Lea $et$ $al.$ (1982) used two pointings of the $Einstein$
SSS to determine that abundances were about solar at the center and stayed so
out to $7^\prime$ from the center. However, Koyama $et$ $al.$ (1991)
found the abundance to drop from $\sim0.5$ solar at the center to only 0.1--0.2
solar beyond $1^\circ$ from the center in a
scanning observation with the $Ginga$ LAC.

Ulmer $et$ $al.$ (1987) found in a $Spartan$ {\it1} observation of the Perseus
cluster (A426) that the iron abundance dropped from 0.8 solar within
$5^\prime$ to 0.4 solar between $6^\prime$ and $20^\prime$ from the center.
A recent reanalysis of the $Spartan$ {\it1} data finds a somewhat steeper
abundance gradient (Kowalski $et$ $al.$ 1993).
Ponman $et$ $al.$
(1990) claim an even steeper abundance gradient in $Spacelab$ {\it2} spectra
of Perseus, with iron practically vanishing beyond $\sim20^\prime$.
Meanwhile, Edge (1990) found no abundance gradient within $\sim20^\prime$ in
$EXOSAT$ data.  Recent $BBXRT$ results by Arnaud $et$ $al.$ (1993) show
no abundance gradient out to $13^\prime$, but the measured abundances
are greater than those indicated by the previously noted spectra from
broad-beam instruments.

For the Coma cluster (A1656), Hughes $et$ $al.$ (1988) and Edge (1990) used
several $EXOSAT$ pointings to deduce that
the iron abundance is roughly constant out to 45$^\prime$.
More recently, Hughes $et$ $al.$ (1993) used a combination of
$Tenma$, $EXOSAT$, and $Ginga$ data to limit the allowed
spatial distribution of iron in the Coma cluster. These
authors found that any gradient in the iron distribution must
be shallow and that, in fact, the total mass of iron in the
cluster out to 2 Mpc can be no less than $\sim$75\% of the
amount estimated assuming a uniform iron distribution.

In $BBXRT$ observations of NGC 1399, the central galaxy in the Fornax
cluster, Serlemitsos $et$ $al.$ (1993) found roughly half-solar abundances out
to $\sim8^\prime$, with weak evidence for an outward decline.  A spatially
unresolved $Ginga$ LAC spectrum of this object indicates somewhat higher,
nearly solar, abundances (Ikebe $et$ $al.$ 1992).

In a continuation of this diverse theme,
recent metallicity determinations from $ROSAT$ PSPC observations of
gas in groups of galaxies have also produced varying results with regard to
overall abundances:
Ponman \& Bertram (1993) found a very low metal abundance
($\sim15\%$ solar) in the compact group HCG 62.
Mulchaey $et$ $al.$ (1993) also found abundances to be small ($\lta20\%$
solar) in the NGC 2300 group.  However,
David $et$ $al.$ (1994) found abundances of $\sim80\%$ solar in the
NGC 5044 group, with no discernible gradient.
The low metallicities in the HCG 62 and NGC 2300 groups are in conflict with
the metallicity---temperature trend found in rich clusters, where cool
clusters have higher abundances than hot clusters do (Hatsukade 1989;
Edge and Stewart 1991).

Summarizing these disparate observations, some
clusters seem to have fairly flat abundance distributions in their inner
regions, while others have centrally enhanced abundances, and some
may have declining abundances in their outskirts. Early results on abundances
in groups show a large scatter, and represent a significant departure from the
metallicity---temperature anti-correlation observed in rich clusters.

These observations raise the question: why are abundances
centrally enhanced in some clusters, but not in others?
In addressing this issue we need to consider
the dominant mechanism(s) by which intracluster
metals are injected into intracluster gas in the first place.
The two mechanisms proposed most often for metal contamination in
intracluster gas are supernovae-driven proto-galactic winds
(Larson \& Dinerstein 1975, Vigroux 1977, De Young 1978)
and ram-pressure stripping of gas from galaxies (Gunn \& Gott 1972,
Norman and Silk 1979, Sarazin 1979, Nepveu 1981).
Whereas proto-galactic winds would inject both metals and
energy into intracluster gas, ram-pressure stripping would inject only metals
(White 1991).   Current observations of X-ray surface brightness profiles
and spectra of intracluster gas indicate that the gas
seems to have a greater specific energy than cluster galaxies,
with the disparity being largest for cool clusters (White 1991).
This suggests the gas has experienced energy deposition in excess of that
associated with cluster collapse, which is indicative of proto-galactic winds
(David $et$ $al.$ 1990, White 1991).
Spatially resolved spectra for many
clusters are required to confirm whether intracluster gas, particularly
in cool clusters, has significantly more specific energy than cluster galaxies.

Observations of the metallicity of distant clusters may also constrain
the source of metals.  The detection of metals in very distant clusters would
suggest a protogalactic wind origin.  If the metals are primarily due to gas
ram-pressure stripped from galaxies, the metallicity may build more slowly
and secularly,
as the central gas density of the cluster increases over time and supplements
galaxy-galaxy stripping encounters.

If galactic winds
are necessary to account for some of the energy content of intracluster gas,
one may expect that the metals borne by the winds would be well-mixed
in the cluster.  This may account for the bulk of the metals in the
intracluster gas, but makes the central abundance enhancement in some
clusters problematic.

There are several possible causes of the centrally enhanced abundances
in some clusters, including:
1) in the context of the galactic wind origin for metals, a population gradient
of wind-blowing galaxies in the cluster may imprint an abundance gradient;
that is, if only early-type galaxies had winds, the fact that the fraction
of early-type galaxies declines outward in a cluster may lead to a
declining metallicity gradient;
2) metal-rich gas may be preferentially stripped from galaxies passing
through the dense gas at cluster centers (Nepveu 1981);
3) metal-rich stellar mass loss in a central dominant galaxy may
accumulate within the galaxy if some fraction of the stellar ejecta does
not participate in its cooling flow;
4) given that the central dominant galaxy in a cluster has the deepest
potential well of any galaxy in the cluster, the metals ejected by its
proto-galactic wind (if it had one) may not escape or be well-mixed,
remaining confined mostly near the cluster center;
5) given the anti-correlation of metallicity and temperature in intracluster
gas, perhaps clusters with central abundance enhancements have ingested
a relatively cool, metal-rich subcluster;
6) heavy ions may have gravitationally
settled from exterior parts of the cluster (Fabian \& Pringle 1977); however,
it is now thought that the timescale for ionic settling to occur is too long
for it to be important (Raphaeli 1978).

The first possibility above, that abundance gradients may
be imprinted by population gradients in wind-blowing galaxies, can be
tested by comparing observed population gradients with the many intracluster
abundance profiles that will be derived in the near future from $ASCA$
satellite data.
Clusters without metallicity gradients in the inner regions may have
particularly low spiral fractions, so any population gradient would be weak
in the inner regions.  The Coma and Perseus clusters are consistent with this.
A2142 provides a possible counterexample, since it is a very hot cluster, so
it is likely to have a small spiral fraction, but the analysis of this paper
suggests it has a substantial central abundance enhancement.
Another problem with this hypothesis is that the
observed temperature---metallicity anti-correlation  for clusters may
provide counterevidence, since cool clusters tend to have smaller fractions
of early-type galaxies, but have the highest metal abundances.

The second possibility, that galaxies are preferentially ram-pressure stripped
in cluster centers, also has some difficulties:
the metallicity of the gas stripped from galaxies must
be at least equal to that of the central regions, which may be near solar.
However, if spiral galaxies are the primary source of stripped gas, the gas
will come largely from the outer parts, where the average metallicity in
extended HI disks is likely to be less than solar.
Furthermore, if the metal-rich gas is largely stripped from early-type
galaxies,
their abundances are turning out to be very sub-solar:  for example,
$BBXRT$ observations of NGC 4472 in the Virgo cluster reveal an iron
abundance of $\sim0.2$ solar (Serlemitsos $et$ $al.$ 1993).
This process is unlikely to produce strong gradients in the cores of
rich clusters like Coma, since the stripping saturates (Hughes $et$ $al.$
1993).

The third and fourth possibilities listed above attribute central abundance
enhancements to the stellar mass loss of the central dominant galaxies.
Even if the central abundances were not 2--3 times greater than the
exterior or average abundances, the total amount
of metals is about two orders of magnitude too large to be generated by current
stellar mass loss rates in the central dominant galaxies; however, it is
possible that the metals had accumulated in the vicinity for a long time,
due to protogalactic winds from or atmospheric stripping of the central
dominant galaxy and neighboring galaxies in the core regions.
The current cooling flows, which extend 100---200 kpc from the cluster
centers, will then consist of this particularly metal-rich gas.
The historical contribution of the central dominant galaxies' mass loss
may be assessed by determining the
physical extent of the abundance-enhanced region.

The fifth possibility mentioned above is that a cool, high-metallicity
subcluster may merge with and settle to the center of a hotter,
lower-metallicity cluster to create a central abundance enhancement.
One may then expect some correlation between the existence of
central abundance enhancements and  signs of a merger.
A496 and A1795 provide counterexamples to this expectation:
A496 appears to have centrally enhanced abundances but has no signs
of a significant recent merger; A1795 has only weak evidence
of centrally enhanced abundances, but the high peculiar velocity of
its central cD is suggestive of a recent merger.
Alternatively, cluster mergers may act in the opposite sense, by washing out
pre-existing metallicity gradients, particularly if the sub-clumps are of
comparable mass, so the gravitational potential is maximally perturbed.
A recent numerical simulation by
Roettiger $et$ $al.$ (1993) suggests that the $dark$ $matter$ component of
subclump material will be thoroughly mixed during the merger, but the
mixture of the $gas$ was not discussed.  A more recent simulation
by Pearce $et$ $al.$ (1994) suggests that gas in cluster cores does not
mix completely during mergers.
This lack of complete gas-mixing is supported by a recent $BBXRT$ observations
of the merging cluster A2256, which suggests that a pre-existing cooling flow
in
one of the two major subclumps has been disrupted, but a region of relatively
cool, X-ray absorbing gas still persists (Miyaji $et$ $al.$ 1993); the merging
system is not yet relaxed, so it is not clear whether this possible cooling
flow
remnant will retain its coherence post-relaxation.  If mergers homogenize
at least the central abundance distributions, then one expects an
anti-correlation between the existence of central abundance enhancements
and signs of significant mergers.  At present, this seems more consistent with
the observations: A496 and Virgo are examples of clusters with
central abundance enhancements but no evidence of a recent disruptive
merger, while Perseus and Coma are clusters with centrally homogeneous
metal distributions and evidence of significant mergers (if the binary
central dominant galaxies in Coma can be interpreted as relics of two
former subclusters).  A1795 may be added to the latter list if the
weakness of its evidence for a central abundance enhancement is emphasized.
Counterexamples to this trend may be provided by
A2142 and A2199, which seem to have centrally enhanced abundances
but also have signs of significant mergers (if binary central dominant galaxies
are again taken to be such a sign in the case of A2142).

Spatially resolved spectra from the $ASCA$ satellite will
be able to provide more accurate assessments of abundance gradients in
intracluster gas.  As $ASCA$ spectra accumulate for many clusters, the
energetics of the gas can be compared to those of the galaxies and the
dark matter
to determine whether protogalactic winds are indicated.  The overall
metallicity and the nature of any metallicity gradients will be
more readily correlated
with other cluster properties, such as spiral fraction,
in order to determine whether the primary metal injection mechanism is
galactic winds or ram-pressure stripping.  Finally, correlating the presence or
absence of central abundance enhancements with such cluster properties as
the local spiral fraction or merger signatures will help determine
whether a separate metal injection mechanism is required to create them.

\bigskip

REW, CSRD, and JPH thank F. Makino and the Institute of Space and Astronautical
Science for their hospitality and scientific support during our stays in Japan.
We also thank K. Arnaud, A. Tennant and O. Day for invaluable help with XSPEC,
and R. Mushotzky for
stimulating discussions.  We especially appreciate the effort by K. Arnaud,
R. Johnstone and D. White in developing the cooling flow model within XSPEC.
JPH thanks M.\ Arnaud and K.\ Yamashita for advice and assistance during
the preliminary reduction of the A2199 $Ginga$ LAC data.  We thank the
referee, Peter Thomas, for helpful suggestions.
REW and CSRD were supported in part by NASA grant NAG 8-228.
Additional support for REW was provided by the NSF and the State of Alabama
via EPSCoR II.  JPH was supported in part by NASA grant NAG 8-181.
This research has made use of data obtained through the High
Energy Astrophysics Science Archive Research Center On-line Service, provided
by the NASA-Goddard Space Flight Center.  This research has also made use
of the NASA/IPAC Extragalactic Database (NED), which is operated by the
Jet Propulsion Laboratory, Caltech, under contract with NASA.

					\vfil\eject
					\pageno=30             %SMALL
%					\pageno=45             %BIG
					\bigskip
					\centerline{
REFERENCES				}
					\medskip
					\leftskip=0.5truein
					\parindent=-0.5truein

Allen, C. W. 1976, $Astrophysical$ $Quantities$, (Athlone: London), p. 31

Allen, S. W., Fabian, A. C., Johnstone, R. M., Nulsen, P. E. J.,
     \& Edge, A. C. 1992, MNRAS, 254, 51

Allen, S. W., Fabian, A. C., Johnstone, R. M., White, D. A., Daines, S. J.,
     Edge, A. C. \& Stewart, G. C. 1993, MNRAS, 262, 901

Arnaud, K. A. \& Fabian, A. C. 1987, preprint

Arnaud, K. A. $et$ $al.$ 1993, preprint

Arnaud M., Lachieze-Rey, M., Rothenflug, R., Yamashita, K. \& Hatsukade, I.
      1991, A\&A, 243, 67

Arnaud, M., Hughes, J. P., Forman, W., Jones, C., Lachieze-Rey, M.,
     Yamashita, K., \& Hatsukade, I. 1992, ApJ, 390, 345

Birkinshaw, M., Hughes, J. P., \& Arnaud, K. A. 1991, ApJ, 379, 466

Canizares, C. R., Markert, T. H, \& Donahue, M. E. 1988, in {\it Cooling Flows
     in Clusters and Galaxies}, ed. A. C. Fabian, Kluwer Academic
     Publishers, Dordrecht, p. 63

Christian, D. J., Swank, J. H., Szymkowiak, A. E., \& White, N. E. 1992,
     $Legacy$, 1, 38

Cowie, L., Hu, E. M., Jenkins, E. B. \& York, D. G. 1983, ApJ, 272, 29 (CHJY)

David, L. P., Arnaud, K. A., Forman, W. \& Jones, C. 1990, ApJ, 356, 32

David, L. P. Forman, W. \& Jones 1994, preprint.

Day, C. S. R., Fabian, A. C., Edge, A. C., Raychaudhury, S. 1991, MNRAS,
     252, 394

De Young, D. 1978, ApJ, 223, 47

Drake, S., Arnaud, K. \& White, N. 1992, $Legacy$, 1, 43

Edge, A. C. 1990, PhD thesis, University of Leicester

Edge, A. C. \& Stewart, G. C. 1991, MNRAS, 252, 414

Edge, A. C., Stewart, G. C., Fabian, A. C. \& Arnaud, K. A. 1990, MNRAS, 245,
     559

Elvis, M., Green, R. F., Bechtold, J., Schmidt, M., Neugebauer, G.,
Soifer, B. T., Mat-thews, K., \& Fabbiano, G. 1986, ApJ, 310, 291

Fabian, A. C. \& Pringle, J. E. 1977, MNRAS, 181, 5p

Gebhardt, K. and Beers, T. C. 1991, ApJ, 383, 72

Gunn, J. E. \& Gott, J. R. III 1972, ApJ, 176, 1

Hatsukade, I. 1989, Ph.D. thesis, Osaka University

Heckman, T. M. 1981, ApJ, 250, L59

Hoessel, J. G. 1980, ApJ, 241, 493

Hughes, J. P., Butcher, J. A., Stewart, G. C., \& Tanaka, Y. 1993 ApJ, 404, 611

Hughes, J. P., Gorenstein, P., \& Fabricant, D. 1988, ApJ, 329, 82

Hughes, J. P. \& Tanaka 1992, ApJ, 398, 62

Ikebe, Y., Ohashi, T., Makishima, K., Tsuru, T., Fabbiano, G., Kim, D.-W.,
     Trinchieri, G., Hatsukade, I., Yamashita, K., \& Kondo, H.
     1992, ApJ, 384, L5

Johnstone, R. M., Fabian, A. C., Edge, A. C. \& Thomas, P. A. 1992,
     MNRAS, 255, 431

Kowalski, M. P., Cruddace, R. G., Snyder, W. A. \& Fritz G. G. 1993, ApJ, 412,
     489

Koyama, K., Takano, S., \& Tawara, Y. 1991, Nature, 350, 135

Larson, R. B. \& Dinerstein, H. L. 1975, PASP, 87, 911

Lea, S. M., Mushotzky, R. F. \& Holt, S. S. 1982, ApJ, 262, 24

Makino, F., \& Astro-C Team 1987, ApLettComm, 25, 223

Masai, K. 1984, ApSpSci, 98, 367

McHardy, I. M., Stewart, G. C., Edge, A. C., Cook, B., Yamashita, K.
     \& Hatsukade, I. 1990, MNRAS, 242, 215

McNamara, B. R. \& O'Connell, R. W. 1992, ApJ, 393, 579

Miyaji, T., Mushotzky, R. F., Loewenstein, M., Serlemitsos, P. J.
     Marshall, F. E., Petre, R., Jahoda, K. M., Boldt, E. A., Holt, S. S.,
     Swank, J., Szymkowiak, A., \& Kelley, R. 1993, ApJ, 419, 66

Morrison, R., \& McCammon, D. 1983, ApJ, 270, 119

Mulchaey, J. S., Davis, D. S., Mushotzky, R. F., \& Burnstein, D. 1993,
     ApJ, 404, L9

Mushotzky, R. F. 1984, Physica Scripta, T7, 157

Mushotzky, R. F. and Szymkowiak, A. E. 1988, in {\it Cooling Flows
     in Clusters and Galaxies}, ed. A. C. Fabian, Kluwer Academic
     Publishers, Dordrecht, p. 53

Nepveu, M. 1981, A\&A, 101, 362

Norman, C. \& Silk, J. 1979, ApJ, 233, L1

Nulsen, P. E. J., Stewart, G. C., Fabian, A. C., Mushotzky, R. F., Holt, S. S.,
	Ku, W., H.-M., Malin, D. F. 1982, MNRAS, 199, 1089

Pearce, F. R., Thomas, P. A. \& Couchman, H. H. M. P. 1994, MNRAS, in press

Ponman T. J. \& Bertram, D. 1993, Nature, 363, 51

Ponman, T. J. $et$ $al.$ 1990, Nature, 347, 450

Raphaeli, Y. 1978, ApJ, 225, 335

Raymond J. C., and Smith, B. W. 1977, ApJS,  35, 419

Reichert, G. A., Mason, K. O., Thorstensen, J. R. \& Bowyer, S. 1982, ApJ,
     260, 437

Roettiger, K., Burns, J., \& Loken, C. 1993, ApJ, 407, L53

Sarazin, C. L. 1979, ApLet, 20, 93

Serlemitsos, P. J., M. Loewenstein, Mushotzky, R. F., Marshall, F. E. \& Petre,
R.
     (1993), ApJ, 413, 518

Stark, A. A., Gammie, C. F., Wilson, R. W., Bally, J., Linke, R. A.,
Heiles, C., \& Hurwitz, M. 1992, ApJS, 79, 77

Stewart, G. C., Fabian, A. C., Jones, C. \& Forman, W. 1984, ApJ, 285, 1

Szymkowiak, A. E. 1986, Ph.D. thesis, University of Maryland
	(NASA Technical Memorandum 86169)

Takano, S., Awaki, H., Koyama, K., Kunieda, H., Tawara, Y., Yamauchi, S.,
     Makishima, K., \& Ohashi, T. 1989, Nature, 340, 289

Thomas, P. A., Fabian, A. C., \& Nulsen, P. E. J. 1987, MNRAS, 228, 973

Turner, M. J. L. $et$ $al.$ 1989, PASJ, 41, 373

Ulmer, M. P. $et$ $al.$ 1987, ApJ, 319, 118

Vigroux, L. 1977, A\&A, 56, 473

Wang, Q. \& Stocke, J. T. 1993, ApJ, 408, 71

White, D. A., Fabian, A. C., Johnstone, R. M., Mushotzky, R. F.,
     and Arnaud, K. A. 1991, MNRAS, 252, 72 (WFJMA)

White, R. E. III 1991, ApJ, 367, 69

White, R. E. III \& Sarazin, C. L. 1987, ApJ, 318, 621

Zabludoff, A. I., Huchra, J. P., \& Geller, M. J. 1990, ApJS, 74, 1 (ZHG)

					\leftskip=0truein
					\parindent=0truein

					\vfil\eject
					\bigskip
					\centerline{
FIGURE CAPTIONS				}
					\medskip
					\leftskip=0.5truein
					\parindent=-0.5truein

{\smc Fig.} 1 -- Two-temperature RS spectral models jointly fit to
       $Einstein$ SSS and $Ginga$ LAC data for A496;
       see line 5 of Table 2 for best-fit parameters.

{\smc Fig.} 2 -- Temperature ($kT$, in keV) $\chi^2$ distributions for
       two-temperature RS models:
	a) A496; b) A1795; c) A2142; d) A2199.
	The lower and upper dashed lines are the 90\% and 99\% confidence
	levels, corresponding to $\Delta\chi^2=2.71$ and 6.63, respectively.

{\smc Fig.} 3 -- $\chi^2$ distributions for the abundance(s) of the hot and
cool
       components in the fits of Fig. 1:
	a) A496; b) A1795 (abundances of the two temperature components
	were set equal); c) A2142; d) A2199 (abundances of the two temperature
	components were set equal).

{\smc Fig.} 4 -- $\chi^2$ distributions for the ``intrinsic" hydrogen column
       density associated with the cooler spectral component of the
       a) A496; b) A1795; c) A2142;
	d) A2199.  Galactic values are also indicated by vertical lines.

{\smc Fig.} 5 -- $N_H$--$T$ contours for isothermal fit to SSS data of A496.

{\smc Fig.} 6 -- $\chi^2$ distributions for abundances of cooling flow (cf)
       and RS (hot) components of joint fits to LAC and SSS spectra.
	a) A496; b) A1795; c) A2142; d) A2199.

{\smc Fig.} 7 -- Two-temperature RS spectral models jointly fit to
       $Einstein$ SSS and $Ginga$ LAC data for A1795; see line 4 of Table 3
       for best-fit parameters.

{\smc Fig.} 8 -- Two-temperature RS spectral models jointly fit to
       $Einstein$ SSS and $Ginga$ LAC data for A2142;
       see line 4 of Table 4 for best-fit parameters.

{\smc Fig.} 9 -- Two-temperature RS spectral models jointly fit to
       $Einstein$ SSS and $Ginga$ LAC data for A2199;
       see line 4 of Table 5 for best-fit parameters.

\bye